\documentclass[twocolumn,english,aps,prx,floatfix,amssymb,superscriptaddress,longbibliography]{revtex4-2}
\usepackage[latin9]{inputenc}
\usepackage{verbatim}
\usepackage{float}
\usepackage{amsmath}
\usepackage{amssymb}
\usepackage{graphicx}
\usepackage{color}
\usepackage{xcolor}
\usepackage{tikz}
\usepackage[normalem]{ulem}

\newcommand{\ket}[1]{\ensuremath{\left| #1 \right>}}

\usepackage{bbold}

\usepackage[bookmarks=false,linkcolor=blue,urlcolor=blue,colorlinks,citecolor=blue]{hyperref}
\makeatletter

\newcommand{\be}{\begin{equation}}
\newcommand{\ee}{\end{equation}}
\newcommand{\bea}{\begin{eqnarray}}
\newcommand{\eea}{\end{eqnarray}}

\begin{document}

\title{Quantum Knizhnik-Zamolodchikov Equations and Integrability of Quantum Field Theories with Time-dependent Interaction Strength}
\author{Parameshwar R. Pasnoori}
\affiliation{Department of Physics, University of Maryland, College Park, MD 20742, United
States of America}
\affiliation{Laboratory for Physical Sciences, 8050 Greenmead Dr, College Park, MD 20740,
United States of America}
\email{pparmesh@umd.edu}
\begin{abstract}

In this paper we consider the problem of solving quantum field theories with time dependent interaction strengths. We show that the recently formulated framework [P. R. Pasnoori, Phys. Rev. B 112, L060409 (2025)], which is a  generalization of the regular Bethe ansatz technique, provides the exact many-body wavefunction. In this framework, the time-dependent Schrodinger equation is reduced to a set of analytic difference equations and matrix difference equations, called the quantum Knizhnik-Zamolodchikov (qKZ) equations. The consistency of the solution gives rise to constraints on the time-dependent interaction strengths. For interaction strengths satisfying these constraints, the system is integrable, and the solution to the qKZ and the analytic difference equations provides the explicit form of the many-body wavefunction that satisfies the time-dependent Schrodinger equation. We provide a concrete example by considering the $SU(2)$ Gross-Neveu model with time dependent interaction strength. Using this framework we solve the model with the most general time-dependent interaction strength and obtain the explicit form of the wave function.
 
\end{abstract}
\maketitle

\section{Introduction}

There has been a resurgence in the study of time-dependent Hamiltonians \cite{PasnooriKondo,Cao_2025,timesimulationclock} due to their applicability in modern experiments ranging from circuit QED \cite{TCSun}, superconducting circuits \cite{timeSCcircuit,vincenzotime} and also cold atom experiments \cite{RAIZENcold}. In addition, due to high coherence times and advanced quantum error correction techniques, it is now possible to simulate time dependent Hamiltonians in digital quantum computers \cite{timetrotter}. Hence, study of exactly solvable or integrable time-dependent Hamiltonians is of high importance.  

Quantum integrability is rooted in Bethe ansatz, which is a powerful mathematical technique that has been very successful in obtaining exact solutions to many-body problems. The Bethe ansatz was originally developed in the form of coordinate Bethe ansatz and was employed to solve the Heisenberg chain \cite{Bethe1931,Hulthen}. It was applied to solve many-body systems in the continuum \cite{LL1,LL2,CNYang,McGuire,GAUDIN,Sutherland,Thacker}, various lattice models \cite{LiebFu} and also vertex models in statistical mechanics \cite{sixV,eightV}.  During these developments, the algebraic structures associated with integrable systems have been found which eventually led to the formulation of the algebraic Bethe ansatz \cite{SklyaninQISM}. This powerful method was successful in obtaining exact solutions to quantum field theories and lattice models with internal degrees of freedom \cite{AndreiLowenstein79,DestriLowenstein,Japaridze,Shastry,Sarkar,FKor,IZERGIN}. It has also been applied to solve many-body quantum impurity models in condensed matter physics \cite{Andrei80,Wiegmann_1981,AndreiLowensein81,parmeshkondo1,parmeshkondo2}. In all the integrable systems, the scattering between particles can be reduced to pair wise scattering processes \cite{ZAMOLODCHIKOVAA}. Each scattering process is associated with an S-matrix and these S-matrices satisfy the quantum Yang-Baxter (QYB) algebra \cite{CNYang}. All models that are integrable by the method of Bethe ansatz have constant interaction strengths and are based on the QYB equation. 

\begin{center}
\begin{figure}
\includegraphics[width=1\columnwidth]{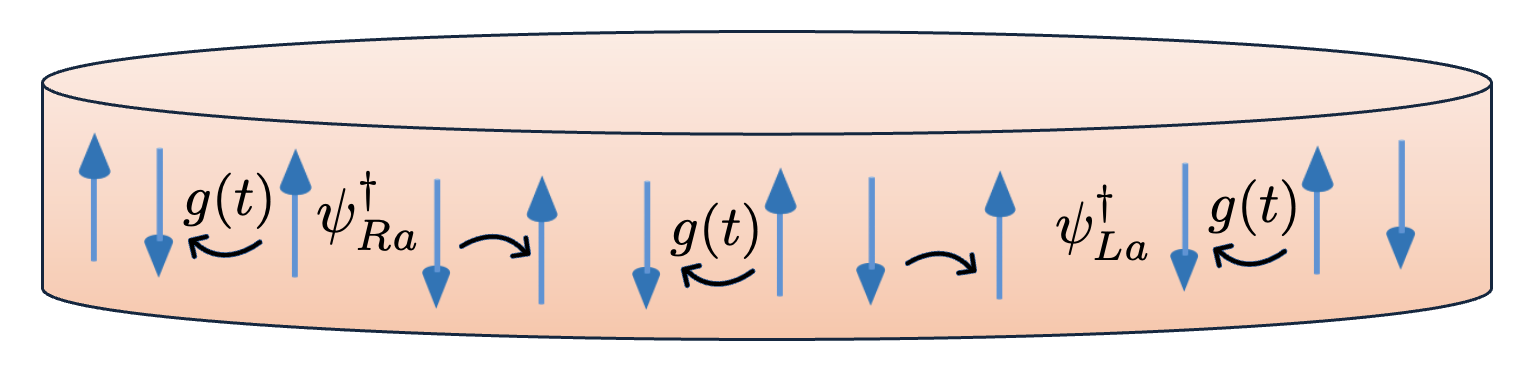}
\caption{Figure depicts electrons (blue arrows) in a quantum wire which forms a loop. The electrons interact with each other through spin exchange interaction with time dependent strength $g(t)$, which is uniform throughout space. }
\label{fig:picture}
\end{figure}
\end{center}

 Recently, a new framework was developed \cite{PasnooriKondo} to analyze Hamiltonians with time-dependent interaction strengths. It was applied to the paradigmatic quantum impurity model, which is the Kondo model with time-dependent interaction strength. Exact many-body wavefunction was constructed, which is the solution to the time-dependent Schrodinger equation.  This framework generalized the standard Bethe ansatz technique and opened a new class of time-dependent Hamiltonians that are based on QYB algebra. Prior to this work, all known integrable Hamiltonians with time-dependent interaction strengths or couplings \cite{Sinytsyn,YUZBASHYAN}
 were based on classical Yang-Baxter (CYB) algebra \cite{RICHARDSON,Gaudin1976,Dukelsky}, and were limited to simple models involving $N\times N$ Hermitian matrices such as multi-level Landau-Zener model \cite{multiLZ}, two level systems interacting with quantized bosonic field such as Tavis-Cummings model \cite{TCSun} and systems based on mean field approximation such as the BCS model \cite{timeBCS} etc. In contrast, the method developed in \cite{PasnooriKondo} is applicable to strongly correlated many-body systems which exhibit strong quantum fluctuations. 

In this work, we shall apply this framework to solve the time-dependent $SU(2)$ Gross-Neveu model, which is a quantum field theory of spin $1/2$ fermions interacting with each other through spin exchange, whose interaction strength is time-dependent (\ref{fig:picture}). The Hamiltonian is given by
 $H_{\text{GN}}\!\!=\!\!\int_{0}^{L}\mathrm{d}x \mathcal{H}_{\rm GN} (x)$, where the Hamiltonian density $\mathcal{H}_{\text{GN}}(x)$ takes the following form
\begin{align}\nonumber
&\mathcal{H}_{\text{GN}}(x)= \sum_{a=\uparrow,\downarrow}\psi^{\dagger}_{L a}(x)i\partial_x \psi^{}_{L a} (x)-\psi^{\dagger}_{R a}(x)i\partial_x \psi^{}_{R a}(x) \\\nonumber&-2g(t)\;\vec{\sigma}_{ab}\cdot\vec{\sigma}_{cd}\sum_{a,b,c,d=\uparrow,\downarrow}\psi^{\dagger}_{Ra}(x)\psi^{\dagger}_{Lc}(x)\psi_{Rb}(x)\psi_{Ld}(x),\\&\text{where,}\;\;\;\vec{\sigma}_{ab}\cdot\vec{\sigma}_{cd}=\left(\sigma^x_{ab}\sigma^x_{cd}+\sigma^y_{ab}\sigma^y_{cd}+\sigma^z_{ab}\sigma^z_{cd}\right).
\label{Hamiltonian}
\end{align} 
Here, the fields $\psi_{L(R) a}(x)$, $a=(\uparrow, \downarrow)$, 
describe left and right moving fermions carrying spin $1/2$. We set their velocity $v_F=1$. The two terms in the first line describe the right and left moving fermions respectively and the third term describes the spin exchange interaction between a left and a right moving fermion as they cross. The interaction strength $g(t)$ is dependent on time and is uniform throughout space.  We apply periodic boundary conditions, where the fermion fields $\psi_{L,R}(x)$ satisfy the following conditions
\be\label{pbc} \psi_{Ra}(0)=  \psi_{Ra}(L),\;\;\psi_{La}(0)=  \psi_{La}(L).\ee
Under these boundary conditions, the total number of left moving fermions $N_L$ and right moving fermions $N_R$ are separately conserved, where
\begin{align}\nonumber &N_L=\sum_a\int_{0}^{L} dx\; \psi^{\dagger}_{La}(x)\psi_{La}(x), \\&N_R=\sum_a\int_{0}^{L}dx\;\psi^{\dagger}_{Ra}(x)\psi_{Ra}(x).\label{number}\end{align}
The total number of fermions in the system $N$ is given by \be N=N_L+N_R. \ee
The Hamiltonian is also $SU(2)$ invariant as it commutes with the total spin operator $\vec{s}$, where
$\vec s_{} =  \int_{0}^{L} dx\; \vec s_{}(x)$ with
\be
\label{Spin}
\vec s_{}(x)= \frac{1}{2} (\psi^{\dagger}_L(x) \vec \sigma \psi^{}_L(x) + \psi^{\dagger}_R(x) \vec \sigma \psi^{}_R(x)).
\ee
The system also exhibits a discrete spin flip symmetry which has the $\mathbb{Z}_2$ group structure, where
\be \tau: \psi_{R\uparrow}(x)\rightarrow\psi_{R\downarrow}(x),\;\;  \psi_{L\uparrow}(x)\rightarrow\psi_{L\downarrow}(x), \;\; \tau^2=1.\ee

In the case of constant interaction strength, the Hamiltonian (\ref{Hamiltonian}) has been solved using Bethe ansatz \cite{AndreiLowenstein79,DestriLowenstein}. In which case, the system exhibits a separation of spin and charge degrees of freedom. Moreover, due to the fermion-fermion interaction, the system exhibits a dynamical generation of mass gap $\Delta$ in the spin sector, where the excitations are called spinons. The charge sector however remains gapless and the charge excitations are called holons. The mass gap stabilizes quasi-long range spin-singlet superconducting order, characterized by 
\begin{align}\nonumber &\langle {\cal O}_s(x) {\cal O}_s(0) \rangle \sim |x|^{-1/2}, \\  {\cal O}_s(x) \propto \: &\psi^{\dagger}_{R\uparrow}(x)\psi^{\dagger}_{L\downarrow}(x) - \psi^{\dagger}_{R\downarrow}(x)\psi^{\dagger}_{L\uparrow}(x).\end{align} 
At high energies, the system exhibits asymptotic freedom \cite{AndreiLowenstein79}. It was show in \cite{PAA1,PAA2} that when twisted boundary conditions are applied, which can be achieved by applying a Zeeman field at the boundaries, the system exhibits a symmetry protected topological (SPT) phase.

Here, we consider this model with time dependent interaction strength  (\ref{Hamiltonian}), and by following the method developed in \cite{PasnooriKondo}, we construct the exact wavefunction which is the solution to the time-dependent Schrodinger equation. The construction of the exact wavefunction involves several steps. Firstly, one identifies certain conserved quantities, which act as quantum numbers that label the wavefunction. The wavefunction is then constructed by ordering the particles in the configuration space, where there exists a unique amplitude corresponding to a specific ordering of all the particles with respect to each other. The amplitudes are related to each other through the action of the particle-particle S-matrices, such that all the amplitudes can be related to any one amplitude. Both the S-matrices and the amplitudes contain a phase part and a spin part. In order for the system to be integrable, these S-matrices should satisfy the quantum Yang-Baxter equation, which guarantees the consistency of the wavefunction and imposes constraints on the time-dependent interaction strength $g(t)$. To determine these amplitudes, one applies periodic boundary conditions on the wavefunction, which gives rise to constraint equations that take the form of matrix difference equations. The solution to the these equations provides the explicit form of the amplitudes and hence also of the exact wavefunction. We show that the matrix difference equations reduce to a set of analytic difference equations and quantum Knizhnik-Zamolodchikov (qKZ) equations. The analytic difference equations govern the phases associated with the S-matrices whereas the qKZ equations govern the spin part. 
In the special case of constant interaction strength, the procedure described above reduces to the regular Bethe ansatz method. Firstly, the amplitudes and the S-matrices are constants, where the phase part of the amplitudes are simple exponential functions. The matrix difference equations reduce to an eigenvalue equation involving a transfer matrix. In the case where the interaction strength is time-dependent, these amplitudes are vector valued functions that depend on the time dependent interaction strength and also on the positions of all the particles. Hence, one can consider the procedure described above as a generalization of the regular Bethe ansatz method.

When the interaction strength is constant, as mentioned above, the regular Bethe ansatz method gives rise to a transfer matrix. This can be diagonalized using the algebraic Bethe ansatz technique \cite{SklyaninQISM}, which yields the eigenstates and the corresponding eigenvalues. In the case of time-dependent interaction strength, as described above, the generalized Bethe ansatz method gives rise to a set of analytic difference equations and the qKZ equations. The solution to these equations provides the explicit form of the amplitudes, and hence also that of the complete wavefunction. This can be achieved by the method of off-shell Bethe ansatz method \cite{rishetikhin1,Babujian_1997}. We consider the most general interaction strength that satisfies the constraint conditions imposed by integrability, and solve the associated analytic difference equations and the qKZ equations, and obtain the explicit form of the exact many-body wavefunction.

The paper is organized as follows. In section {\ref{sec:ansatzwf}} we present the ansatz for the wavefunction that satisfies the time-dependent Schrodinger equation and provide the constraint conditions on the interaction strength. We then apply periodic boundary conditions on the wavefunction, which gives rise to matrix difference equations. In section (\ref{sec:wavefunction}), we work with the most general interaction strength that satisfies the constraint conditions and show that the matrix difference equations take the form of qKZ equations. We then present the solution to these equations, thus obtaining the explicit form of the many-body wavefunction. In section (\ref{sec:discussion}), we conclude and discuss future prospects.

\section{The ansatz wavefunction}
\label{sec:ansatzwf}
As mentioned above, the number of left and right moving fermions are separately conserved (\ref{number}). Hence, one can construct the ansatz wavefunction which consists of $N_R$ number of right moving fermions and $N_L$ number of left moving fermions, which we denote by $\ket{N_L,N_R}$. This wavefunction satisfies the following time-dependent Schrodinger equation 
\be i\partial_t\ket{N_L,N_R}=H_{\text{GN}} \ket{N_L,N_R},\label{SE}\ee
where $H_{\text{GN}}$ is the Hamiltonian (\ref{Hamiltonian}).

\subsection{One particle case}
Let us first consider the most simplest case of one particle $N=1$. There are two possible wavefunctions corresponding to the choices: $N_R=1,N_L=0$ and $N_L=0,N_R=1$, which we label by $\ket{0,1}$ and $\ket{1,0}$ respectively.  
We have 
\begin{align}
    &\ket{0,1}=\sum_a\int_0^L dx \; \psi^{\dagger}_{Ra}(x)F_{Ra}(x,t)\ket{0},\\&\ket{1,0}=\sum_a\int_0^L dx \; \psi^{\dagger}_{La}(x)F_{La}(x,t)\ket{0}.\end{align}
Here, $a=\uparrow\downarrow$ denotes the spin of the fermions. Using these in the Schrodinger equation (\ref{SE}), we obtain the following one particle Schrodinger equations
\begin{align}\label{1pse}i(\partial_t+\partial_x)F_{Ra}(x,t)=0,\;\; i(\partial_t-\partial_x)F_{La}(x,t)=0.\end{align}
To solve the above equations, we consider the following ansatz
\begin{align}\nonumber
&F_{Ra}(x,t)=f_{Ra}(z)\;\theta(x)\theta(L-x),\\& F_{La}(x,t)=f_{La}(\bar{z})\;\theta(x)\theta(L-x).\label{1pansatz}\end{align}
Here, we have used the notation $z=x-t, \bar{z}=x+t$. $\theta(x)$ is the Heaviside step function with the convention $\theta(x)=1$ for $x>0$, $\theta(x)=0$ for $x<0$  and $\theta(0)=1/2$. It is understood that $f_{Ra}(z)$ and $f_{La}(\bar{z})$ are vectors in the spin space of the particle. The boundary conditions (\ref{pbc}) give rise to the following conditions on the amplitudes
\be f_{Ra}(z)=f_{Ra}(z+L),\;\;
f_{La}(\bar{z})=f_{La}(\bar{z}+L).\ee
One can see that the above ansatz (\ref{1pansatz}) satisfies the Schrodinger equations (\ref{1pse}) trivially. 

Let us compare this with the standard Bethe ansatz procedure that one applies when the interaction strength is constant $g(t)=g$. In this case, the functions $f_{Ra}(z)$ and $f_{La}(\bar{z})$ can be expressed in terms of simple exponential functions
\begin{align}f_{Ra}(z)=e^{ikz}A_{Ra},\;\; f_{La}(z)=e^{-ik\bar{z}}A_{La},\label{1conststrength}\end{align}
where $k$ is the momentum, which is also the energy since we have set $v_F=1$, and $A_{Ra}$,$A_{La}$ are amplitudes which do not depend on $z$ and $\bar{z}$.

\subsection{Two particle case and the S-matrix}
Now let us consider the case of two particles $N=2$. There are three possibilities: 
\begin{align} 1)\;N_L=2,\;\;N_R=0: \;\;\ket{2,0},\\ 2)\;N_L=0,\;\;N_R=2: \;\; \ket{0,2} ,\\ 3)\;N_L=1,\;\;N_R=1:\;\; \ket{1,1}.\end{align} 
The first two sectors which correspond to both the particles being either left movers or right movers is again trivial, as the two particles do not interact with each other. The two particle wavefunction is just the direct product of one particle wavefunctions discussed above. Now let us consider the last sector where one particle is a left mover whereas the other is a right mover. This sector is non trivial since the two particles interact with each other through the spin exchange interaction in the Hamiltonian. We have
\begin{align}\ket{1,1}=\prod_{i=1}^2\int_0^L \hspace{-1mm} dx_i \;\psi^{\dagger}_{Ra}(x_1)\psi^{\dagger}_{Lb}(x_2) \mathcal{A}F^{RL}_{ab}(x_1,x_2,t)\ket{0}.\end{align}
Here the superscripts denote the chirality and the subscripts denote the spin of the fermions. $\mathcal{A}$ is the anti-symmetrization symbol which when acting on $F^{RL}_{ab}(x_1,x_2,t)$ exchanges $x_1\leftrightarrow x_2$, $R\leftrightarrow L$ and $a\leftrightarrow b$. Using the above expression in the Schrodinger equation (\ref{SE}), one obtains the following two particle Schrodinger equation
\begin{align}\nonumber
    -i(\partial_t+\partial_{x_1}-\partial_{x_2})&\mathcal{A}F^{RL}_{ab}(x_1,x_2,t)+g(t)\delta(x_1-x_2)\\&\times(\vec{\sigma}_{ac}\cdot\vec{\sigma}_{bd})\;\mathcal{A}F^{RL}_{cd}(x_1,x_2,t)=0\label{se2p}.
\end{align}
Since a left moving and a right moving fermion interact with each other, the ordering of the particles is important. We take the following ansatz for the wavefunction $F^{RL}_{ab}(x_1,x_2,t)$:
\begin{align}\nonumber
  &F^{RL}_{ab}(x_1,x_2,t)= \theta(x_1)\theta(L-x_1)\theta(x_2)\theta(L-x_2)\times\\&  \big(f^{RL,12}_{ab}(z_1,\bar{z}_2) \theta(x_2-x_1)+f^{RL,21}_{ab}(z_1,\bar{z}_2)\theta(x_1-x_2) \big).\label{2pwf}
\end{align}
The ordering of the particles is denoted by the superscripts. For example,  $f^{RL,12}_{ab}(z_1,\bar{z}_2)$ corresponds to the amplitude in which the particle 1, which is a right mover, is on the left side of particle 2 which is a left mover. Using the above ansatz (\ref{2pwf}) in the two particle Schrodinger equation (\ref{se2p}), one obtains the following relation between the amplitudes
\begin{align}
f^{RL,21}_{ab}(z_1,\bar{z}_2)=S^{12}_{ac,bd}(z_1,\bar{z}_2)f^{RL,12}_{cd}(z_1,\bar{z}_2),\label{2prel}\end{align}
where $S_{12}(z_1,\bar{z}_2)$ is the two particle S-matrix which is given by
\begin{align}\nonumber
   &S^{12}_{ac,bd}(z_1,\bar{z}_2)=e^{i\phi(\bar{z}_2-z_1)}\frac{ic(\bar{z}_2-z_1)I^{12}_{ac,bd}+P^{12}_{ac,bd}}{ic(\bar{z}_2-z_1)+1},\\\nonumber&c(x)=\frac{1}{2g(x/2)}\left(1-\frac{3(g(x/2))^2}{4}\right),\\&e^{i\phi(x)}=\frac{2ig(x/2)-1+\frac{3(g(x/2))^2}{4}}{ig(x/2)-\left(1+\frac{3(g(x/2))^2}{4}\right)}.\label{smatlr}
   \end{align}
Here, $I^{12}_{ac,bd}$ is the identity and $P^{12}_{ac,bd}$ is the permutation operator. The superscripts in $I^{12}_{ac,bd}, P^{12}_{ac,bd}$ and in $S^{12}_{ac,bd}(z_1,\bar{z}_2)$ denote that they act in the spin spaces of particles 1 and 2 which are represented by the subscripts. Note that the relation between the interaction strength $g(t)$ and the function $c(t)$ depends on the regularization of the $\theta(x)$. The relation between these functions provided above (\ref{smatlr}) arise as a result of our definition of $\theta(x)$ (see below Eq (\ref{1pansatz})). Use of a different regularization of $\theta(x)$ yields a different relation between these functions. However, we note that there is no discrepancy. In the limit where the interaction strengths are small, expressions arising from different regularization schemes coincide, which corresponds to the universal regime \cite{Pasnoori2025RGint}. 

We see that the interaction between the particles relates the two amplitudes in the two particle wave function (\ref{2pwf}) such that there exists one free amplitude. We may choose this to be $f^{RL,12}_{ab}(z_1,\bar{z}_2)$. 

Let us now apply the periodic boundary conditions (\ref{pbc}) on the two particle wavefunction (\ref{2pwf}). This results in the following relations between the amplitudes
\begin{align}\nonumber
&f^{RL,12}_{ab}(z_1,\bar{z}_2)=f^{RL,21}_{ab}(z_1+L,\bar{z}_2),\\&
f^{RL,21}_{ab}(z_1,\bar{z}_2)=f^{RL,12}_{ab}(z_1,\bar{z}_2+L).\label{2ppbc}
\end{align}
Using the relations (\ref{2ppbc}) in (\ref{2prel}), we obtain
\begin{align}
f^{RL,12}_{ab}(z_1,\bar{z}_2+L)=S^{12}_{ac,bd}(z_1,\bar{z}_2)f^{RL,12}_{cd}(z_1,\bar{z}_2).\label{diffeq}\end{align}
Hence, we find that applying periodic boundary conditions (\ref{pbc}) on the two particle wavefunction (\ref{2ppbc}) imposes constraints on the free amplitude (\ref{diffeq}), which is a matrix difference equation. One can solve this difference equation, and determine the amplitude  $f^{RL,12}_{ab}(z_1,\bar{z}_2)$. One can then use the relation (\ref{2prel}) to determine the other amplitude, and hence the exact form of the two particle wavefunction (\ref{2pwf}).

\subsection{Three particle case and the Yang-Baxter equation}
Let us now consider the case of three particles $N=3$. There exist four possibilities: 
\begin{align}\nonumber1)\;N_L=3,\;\;N_R=0: \;\;\ket{3,0},\\ \nonumber2)\;N_L=0,\;\;N_R=3: \;\; \ket{0,3} ,\\\nonumber 3)\;N_L=2,\;\;N_R=1:\;\; \ket{1,2},\\4)\;N_L=1,\;\;N_R=2:\;\; \ket{2,1}.\label{3pcases}\end{align}
Just as in the case of two particles, the first two sectors are trivial where the three particle wavefunction is a direct product of one particle wavefunctions. The third and the fourth sectors are non-trivial, which we discuss below. Let us first consider the third sector corresponding to two left moving particles and one right moving particle. We have
\begin{align}\nonumber\ket{1,2}=\hspace{-1mm}\prod_{i=1}^3 \int_0^L \hspace{-1mm} dx_i \: \psi^{\dagger}_{Ra}(x_1)\psi^{\dagger}_{Lb}(x_2)\psi^{\dagger}_{Lc}(x_3) \\\times\mathcal{A} F^{RLL}_{abc}(x_1,x_2,x_3,t)\ket{0}.
    \end{align}
Using this in the Schrodinger equation, we obtain the following three particle Schrodinger equation
\begin{align}\nonumber-i(\partial_t+\partial_{x_1}-\partial_{x_2}-\partial_{x_3})\mathcal{A}F^{RLL}_{abc}(x_1,x_2,x_3,t)\\\nonumber+g(t)\left(\delta(x_1-x_2)\vec{\sigma}_{al}\cdot\vec{\sigma}_{bm}\; I_{cn}
+\delta(x_1-x_3)I_{bm}\;\vec{\sigma}_{al}\cdot\vec{\sigma}_{cn}\right)&\\\times\mathcal{A}F^{RLL}_{lmn}(x_1,x_2,x_3,t)=0.\label{se3p}\end{align}
Due to the interactions in the system, just as in the two particle case, the ordering of the left and the right moving particles is important. In addition, we find that the ordering of two left moving particles with respect to each other is also important to have a consistent wavefunction. The ansatz for the three particle wavefunction $\mathcal{A}F^{RLL}_{abc}(x_1,x_2,x_3,t)$ takes the following form
\begin{widetext}
\begin{align}\nonumber
F^{RLL}_{abc}(x_1,x_2,x_3,t)=\theta(x_1)\theta(L-x_1)\theta(x_2)\theta(L-x_2)\theta(x_3)\theta(L-x_3)\times\\\nonumber \big(f^{RLL,123}_{abc}(z_1,\bar{z}_2,\bar{z}_3)\theta(x_2-x_1)\theta(x_3-x_2)+
f^{RLL,132}_{abc}(z_1,\bar{z}_2,\bar{z}_3)\theta(x_2-x_3)\theta(x_3-x_1)+\\\nonumber f^{RLL,213}_{abc}(z_1,\bar{z}_2,\bar{z}_3)\theta(x_1-x_2)\theta(x_3-x_1)+
f^{RLL,312}_{abc}(z_1,\bar{z}_2,\bar{z}_3)\theta(x_2-x_1)\theta(x_1-x_3)+\\f^{RLL,231}_{abc}(z_1,\bar{z}_2,\bar{z}_3)\theta(x_1-x_3)\theta(x_3-x_2)+
f^{RLL,321}_{abc}(z_1,\bar{z}_2,\bar{z}_3)\theta(x_2-x_3)\theta(x_1-x_2)\big).\label{3pwf}\end{align}
\end{widetext}
Using this in the three particle Schrodinger equation (\ref{se3p}), we obtain the following relations (from here on we suppress the spin indices, unless necessary)
\begin{align}\nonumber
f^{RLL,213}(z_1,\bar{z}_2,\bar{z}_3)=S^{12}(z_1,\bar{z}_2)f^{RLL,123}(z_1,\bar{z}_2,\bar{z}_3),\\\nonumber f^{RLL,231}(z_1,\bar{z}_2,\bar{z}_3)=S^{13}(z_1,\bar{z}_3)f^{RLL,213}(z_1,\bar{z}_2,\bar{z}_3),\\\nonumber f^{RLL,312}(z_1,\bar{z}_2,\bar{z}_3)=S^{13}(z_1,\bar{z}_3)f^{RLL,132}(z_1,\bar{z}_2,\bar{z}_3),\\f^{RLL,321}(z_1,\bar{z}_2,\bar{z}_3)=S^{12}(z_1,\bar{z}_2)f^{RLL,312}(z_1,\bar{z}_2,\bar{z}_3).\label{3prelLR}
\end{align}
Here, the S-matrices $S^{12}(z_1,\bar{z}_2)$ and $S^{13}(z_1,\bar{z}_3)$ take the same form as in the two particle case (\ref{smatlr}). 

Note that the relation between the amplitudes which differ by the ordering of the particles with the same chirality is not fixed by the Hamiltonian. In the current case this corresponds to the relation between the amplitudes $f^{RLL,123}(z_1,\bar{z}_2,\bar{z}_3)$ and $f^{RLL,132}(z_1,\bar{z}_2,\bar{z}_3)$ and also between the amplitudes $f^{RLL,231}(z_1,\bar{z}_2,\bar{z}_3)$ and $f^{RLL,321}(z_1,\bar{z}_2,\bar{z}_3)$. This occurs due to the linear dispersion, and also occurs in the case when the interaction strength is constant. Integrability requires one to choose a specific relation between such amplitudes for the wavefunction to be consistent, which we discuss below. Note that up until now, the interaction strength $g(t)$ is an arbitrary function of time. Choosing a consistent relation between the above mentioned amplitudes imposes constraints on the interaction strength $g(t)$. Integrability requires that  $f(t)$ (\ref{smatlr}) is a linear function.
\begin{align} c(t)= \alpha t+\beta , \;\; \alpha,\beta\in \mathcal{C}\;  (\text{constants}).\label{intconst}
\end{align}
which corresponds to the interaction strength taking the following form
\begin{align}
    g(t)=\frac{2}{3}\left(-2(2\alpha t+\beta)\pm \left(4\left(2\alpha t+\beta\right)^2+3\right)^{1/2}\right).\label{intstrength}
\end{align}
Here we stress that the above expression of the interaction strength is non-universal. As mentioned above, it depends on the specific regularization of $\theta(x)$ used. Consistency requires that the S-matrix between the amplitudes is given by
\begin{align}\nonumber
f^{RLL,132}(z_1,\bar{z}_2,\bar{z}_3)=
S'^{23}(\bar{z}_2,\bar{z}_3)f^{RLL,123}(z_1,\bar{z}_2,\bar{z}_3)\\
f^{RLL,321}(z_1,\bar{z}_2,\bar{z}_3)=
S'^{23}(\bar{z}_2,\bar{z}_3)f^{RLL,231}(z_1,\bar{z}_2,\bar{z}_3),\label{3prelLL}
\end{align}
where $S^{23}(\bar{z}_2,\bar{z}_3)$ takes a similar form as that of the two particle S-matrix (\ref{smatlr}), but now with $f(t)$ being a linear function:
\begin{align}&S'^{23}_{ac,bd}(\bar{z}_2,\bar{z}_3)=\frac{i\alpha(\bar{z}_3-\bar{z}_2)I^{32}_{ac,bd}+P^{32}_{ac,bd}}{i\alpha(\bar{z}_3-\bar{z}_2)+1}.\label{smatll}\end{align}
Note that the case where $\alpha$ or $\beta$ are complex gives rise to a non-Hermitian Hamiltonian, which is also interesting. The S-matrices between particles of different chiralities $S^{12}(z_1,\bar{z}_2)$, $S^{13}(z_1,\bar{z}_3)$ and that between the particles of same chirality $S^{32}(\bar{z}_2,\bar{z}_3)$ satisfy the Yang-Baxter equation
\begin{align}\nonumber
S^{12}(z_1,\bar{z}_2)S^{13}(z_1,\bar{z}_3)S'^{23}(\bar{z}_2,\bar{z}_3)=\\S'^{23}(\bar{z}_2,\bar{z}_3)S^{13}(z_1,\bar{z}_3)S^{12}(z_1,\bar{z}_2),\label{YB}\end{align}
which guarantees the consistency of the three particle wavefunction (\ref{3pwf}). Note that the linearity of $f(t)$ is crucial for the S-matrices to satisfy the Yang-Baxter equation.

Using the relations (\ref{3prelLR}) and (\ref{3prelLL}), one can express all the amplitudes in the three particle wavefunction (\ref{3pwf}) in terms of one amplitude of our choosing. We may choose this to be $f^{RLL,123}_{abc}(z_1,\bar{z}_2,\bar{z}_3)$. Hence, we find that, just as in the two particle case, there exists one free amplitude. To find the exact form of the wavefunction, we have to impose periodic boundary conditions (\ref{pbc}) on the three particle wave-function (\ref{3pwf}). This results in the following relations between the amplitudes
\begin{align}\nonumber
 f^{RLL,123}(z_1,\bar{z}_2,\bar{z}_3)=f^{RLL,231}(z_1+L,\bar{z}_2,\bar{z}_3),\\\nonumber f^{RLL,213}(z_1,\bar{z}_2,\bar{z}_3)=f^{RLL,132}(z_1,\bar{z}_2+L,\bar{z}_3),\\\nonumber f^{RLL,312}(z_1,\bar{z}_2,\bar{z}_3)=f^{RLL,123}(z_1,\bar{z}_2,\bar{z}_3+L),\\\nonumber f^{RLL,231}(z_1,\bar{z}_2,\bar{z}_3)=f^{RLL,312}(z_1,\bar{z}_2+L,\bar{z}_3), \\\nonumber f^{RLL,321}(z_1,\bar{z}_2,\bar{z}_3)=f^{RLL,213}(z_1,\bar{z}_2,\bar{z}_3+L),\\f^{RLL,132}(z_1,\bar{z}_2,\bar{z}_3)=f^{RLL,321}(z_1+L,\bar{z}_2,\bar{z}_3).\label{3ppbc} 
 \end{align}
Using the relations (\ref{3prelLR}), (\ref{3prelLL}) and (\ref{3ppbc}), we obtain the following relation constraining the amplitude $f^{RLL,123}_{abc}(z_1,\bar{z}_2,\bar{z}_3)$:

\begin{align}
&f^{RLL,123}(z_1-L,\bar{z}_2,\bar{z}_3)=Z_1(z_1,\bar{z}_2,\bar{z}_3)f^{RLL,123}(z_1,\bar{z}_2,\bar{z}_3),\label{3pme1}\end{align}
where the \textit{transport operator}
$Z_1(z_1,\bar{z}_2,\bar{z}_3)$ transports the particle 1 across the entire system once. It takes the following form
\begin{align}Z_1(z_1,\bar{z}_2,\bar{z}_3)=S^{13}(z_1,\bar{z}_3)S^{12}(z_1,\bar{z}_2).    
\end{align}
Similarly, there exist transport operators $Z_2(z_1,\bar{z}_2,\bar{z}_3)$ and $Z_3(z_1,\bar{z}_2,\bar{z}_3)$ which transport particles 2 and 3 respectively around the system
\begin{align}
    \nonumber&f^{RLL,123}(z_1,\bar{z}_2-L,\bar{z}_3)=Z_1(z_1,\bar{z}_2,\bar{z}_3)f^{RLL,123}(z_2,\bar{z}_2,\bar{z}_3),\\& f^{RLL,123}(z_1,\bar{z}_2,\bar{z}_3-L)=Z_1(z_1,\bar{z}_2,\bar{z}_3)f^{RLL,123}(z_2,\bar{z}_2,\bar{z}_3).\label{3pme2}
\end{align}
These transport operators take the following form
\begin{align}\nonumber&Z_2(z_1,\bar{z}_2,\bar{z}_3)=S^{21}(\bar{z}_2,z_1+L,)S'^{23}(\bar{z}_2,\bar{z}_3),\\&Z_3(z_1,\bar{z}_2,\bar{z}_3)=S'^{32}(\bar{z}_3,\bar{z}_2+L)S^{31}(\bar{z}_3,z_1+L).\end{align}
The equations (\ref{3pme1}) and (\ref{3pme2}) form a system of matrix difference equations. We shall see later in the next section that they are related to quantum Knizhnik-Zamolodchikov equations. Using the Yang-Baxter equation (\ref{YB}), one can show that these transport operators satisfy the following relations
\begin{widetext}
\begin{align}\nonumber&Z_{1}(z_1,\bar{z}_2-L,\bar{z}_3)Z_2(z_1,\bar{z}_2,\bar{z}_3)=Z_2(z_1-L,\bar{z}_2,\bar{z}_3)Z_1(z_1,\bar{z}_2,\bar{z}_3),\\\nonumber&Z_{1}(z_1,\bar{z}_2,\bar{z}_3-L)Z_3(z_1,\bar{z}_2,\bar{z}_3)=Z_3(z_1-L,\bar{z}_2,\bar{z}_3)Z_1(z_1,\bar{z}_2,\bar{z}_3),\\&Z_{2}(z_1,\bar{z}_2,\bar{z}_3-L)Z_3(z_1,\bar{z}_2,\bar{z}_3)=Z_3(z_1,\bar{z}_2-L,\bar{z}_3)Z_2(z_1,\bar{z}_2,\bar{z}_3).\label{comm3p}\end{align}
\end{widetext}
Given a certain interaction strength $g(t)$, which satisfies (\ref{intconst}), one should solve the set of matrix difference equations described above to obtain the amplitude $f^{RLL,123}_{abc}(z_1,\bar{z}_2,\bar{z}_3)$. One can then solve for the rest of the amplitudes in the three particle wavefunction (\ref{3pwf}) using the relations (\ref{3prelLL}), (\ref{3prelLR}) and obtain the explicit form of the wavefunction. 

Similarly, in the sector corresponding to two right moving particles and one left moving particle (\ref{3pcases}), one can construct the wavefunction exactly as in the above case. One finds that the constraints on the interaction strength are exactly the same as (\ref{intconst}), and the associated matrix difference equations take the same form as above. We shall postpone the discussion of the solution of these matrix difference equations to the later sections and consider the most general case of $N$ particles.

\subsection{N particle case and the quantum Knizhnik-Zamolodchikov equations}
Let us now consider the most general case of $N$ particles. Since the number of left movers $N_L$ and right movers $N_R$ are separately conserved under periodic boundary conditions, we have $N+1$ possible sectors corresponding to different values of $N_L$ and $N_R$, where $N_L+N_R=N$. Below, we shall construct the exact many-body wavefunction for general values of $N_L$ and $N_R$. We have
\begin{align}\nonumber
\ket{N_L,N_R}=\prod_{j=N_{L}+1}^{N}\prod_{k=1}^{N_L}\int_{0}^L\hspace{-2.5mm}dx_j\hspace{-1mm}\int_{0}^L\hspace{-2.5mm}dx_k \; \psi^{\dagger}_{R\sigma_j}(x_j)\psi^{\dagger}_{L\sigma_k}(x_k)\\\times\mathcal{A}F^{N...1,\chi_1...\chi_N}_{\sigma_1...\sigma_N}(x_1,...,x_N,t)\ket{0}.\label{npwf1}
\end{align}
Here, $\sigma_1...\sigma_N\equiv\{\sigma_i\}$ denote the spin and $\chi_1...\chi_N\equiv\{\chi_i\}$ denote the chiralities of the particles. Using this in the Schrodinger equation (\ref{SE}), we obtain the following $N$ particle Schrodinger equation 

\begin{align}\nonumber
 &-i(\partial_t+\hspace{-3mm}\sum_{j=N_L+1}^{N}\hspace{-3mm}\partial_{x_j}\hspace{-1mm}-\hspace{-1mm}\sum_{k=1}^{N_L}\partial_{x_k})\mathcal{A}F^{N...1,\{\chi_i\}}_{\sigma_1...\sigma_N}(x_1,...,x_N,t)\hspace{-0.5mm}+\hspace{-0.5mm}g(t)\times\\&\sum_{\substack{j=N_L+1\\k=1}}^{\substack{j=N\\k=N_L}}\hspace{-3mm}\delta(x_j-x_k)\vec{\sigma}_{\sigma_j\sigma'_j}\hspace{-1mm}\cdot\hspace{-0.5mm}\vec{\sigma}_{\sigma_k\sigma'_k}\mathcal{A}F^{N...1,\{\chi_i\}}_{\sigma_1..\sigma'_j\sigma'_k..\sigma_N}\hspace{-0.4mm}(x_1,...,x_N,t)\hspace{-0.7mm}=\hspace{-0.5mm}0.\label{senp}
\end{align}
From here on we use the convention that $F^{N...1,\{\chi_i\}}_{\sigma_1...\sigma_N}(x_1,...,x_N)$ is a vector in the spin space of the particles, as opposed to an amplitude as represented in (\ref{npwf1}). That is  $F^{N...1\{\chi_i\}}_{\sigma_1...\sigma_N}(x_1,...,x_N)\equiv F^{N...1,\{\chi_i\}}_{\sigma_1...\sigma_N}(x_1,...,x_N)\ket{\{\sigma_i\}}$. The ansatz for $F^{N...1,\{\chi_i\}}_{\sigma_1...\sigma_N}(x_1,...,x_N)$ takes the following form

\begin{align} 
F^{N..1,\{\chi_i\}}_{\sigma_1...\sigma_N}\hspace{-0.3mm}(x_1,..,x_N,t)
\hspace{-1mm}= \hspace{-1.3mm}\sum_Q \theta(\{x_{Q(j)}\})  f^{Q,\{\chi_i\}}_{\sigma_1...\sigma_N}(\bar{z}_1,..,z_N).\label{npwf2}
\end{align}
In this wavefunction, without losing generality, we have chosen the particles $i=1,...,N_L$ to be left movers, and $i=N_L+1,...,N$ to be right movers. In the above expression, $Q$ denotes a permutation of the position orderings of particles and  $\theta(\{x_{Q(j)}\})$ is the Heaviside function that vanishes unless $x_{Q(1)} \le \dots \le x_{Q(N)}$. Here $f^{Q}_{\sigma_1...\sigma_N} (\bar{z}_1,...,z_N)$ is the amplitude corresponding to the ordering of the particles denoted by $Q$. The amplitudes that differ by the ordering of the particles with different chiralities are related by the S-matrix (\ref{smatlr}), just as in the two particle case
\begin{align}
f^{...kj...,\{\chi_i\}} (\bar{z}_1,...,z_N)=S^{jk}(z_j,\bar{z}_k)f^{...jk...,\{\chi_i\}} (\bar{z}_1,...,z_N).\label{rel1N}
\end{align}
Here $\chi_j=+,\chi_k=-$ and $``..."$ in the first superscript on both side of the above equation corresponds to any specific ordering of the rest of the particles. In addition to this, just as in the three particle case, the amplitudes that differ by the ordering of the particles with the same chirality are related by an S-matrix. In the case of two left moving particles, we have
\begin{align}
f^{...kj...,\{\chi_i\}}(\bar{z}_1,...,z_N)=S'^{jk}(\bar{z}_j,\bar{z}_k)f^{...jk...,\{\chi_i\}}(\bar{z}_1,...,z_N),\label{rel2N}
\end{align}
where $\chi_{j,k}=-$, and in the case of two right moving particles, we have
\begin{align}
f^{...kj...,\{\chi_i\}}(\bar{z}_1,...,z_N)=S'^{jk}(z_j,z_k)f^{...jk...,\{\chi_i\}}(\bar{z}_1,...,z_N),\label{rel3N}
\end{align}
where $\chi_{j,k}=+$. Just as before, $``..."$ in the first superscript on both side of the above two equations corresponds to any specific ordering of the rest of the particles. These S-matrices satisfy the Yang-Baxter equation. For one right moving particle $i$ and two left moving particles $j$ and $k$, we have
\begin{align}\nonumber
S^{ij}(z_i,\bar{z}_j)S^{ik}(z_i,\bar{z}_k)S'^{jk}(\bar{z}_j,\bar{z}_k)=\\S'^{jk}(\bar{z}_j,\bar{z}_k)S^{ik}(z_i,\bar{z}_k)S^{ij}(z_i,\bar{z}_j).\label{YBn1}\end{align}
Similarly, for two right moving particles $i$ and $j$ and one left moving particle $k$, we have
\begin{align}\nonumber S^{jk}(z_j,\bar{z}_k)S^{ik}(z_i,\bar{z}_k)S'^{ij}(z_i,z_j)=\\S'^{ij}(z_i,z_j)S^{ik}(z_i,\bar{z}_k)S^{jk}(z_j,\bar{z}_k).\label{YBn2}\end{align}
In addition to this, the S-matrices corresponding to the exchange of the particles with the same chirality also satisfy the Yang-Baxter equation. For three right moving particles, we have
\begin{align}\nonumber
S'^{ij}(z_i,z_j)S'^{ik}(z_i,z_k)S'^{jk}(z_j,z_k)=\\S'^{jk}(z_j,z_k)S'^{ik}(z_i,z_k)S'^{ij}(z_i,z_j).\label{YBn3}\end{align}
Similar expression exists for three left moving particles, which can be obtained by applying the transformation $z_{i,j,k}\rightarrow\bar{z}_{i,j,k}$ to the above equation (\ref{YBn3}). Using the relations (\ref{rel1N}) and(\ref{rel2N}), one can express all the amplitudes in the $N$ particle wavefunction (\ref{npwf2}) in terms of one amplitude of our choosing. We  may choose this to be $f^{N...1,\{\chi_j\}}_{\sigma_1...\sigma_N}(\bar{z}_1,...,z_N)$. To obtain the explicit form of the wavefunction, this free amplitude needs to be determined, which can be achieved by applying periodic boundary conditions (\ref{pbc}) on the $N$ particle wavefunction (\ref{npwf2}). Applying periodic boundary conditions yields the following relation
\begin{align}
f^{j...,\{\chi_i\}}(\bar{z}_1,..,z_j,..,z_N)=f^{...j,\{\chi_i\}} (\bar{z}_1,..,z_j+L,..,z_N).\label{nppbc}\end{align}
Here $j$ is a right moving particle. Similar expression exists for a left moving particle, which can be obtained by applying the transformation $z_j\rightarrow\bar{z}_j$ to the above equation (\ref{nppbc}). In the above equation, $``..."$ in the first superscript corresponds to any particular ordering of the rest of the particles, which is same on both sides of the equation. Using the relations (\ref{rel1N}), (\ref{rel2N}), (\ref{rel3N}) and (\ref{nppbc}) we obtain the following constraint equations on the amplitude 

\begin{align}\nonumber
f^{N...1,\{\chi_i\}}(\bar{z}_1,..,z_j-L,..,z_N)=Z_{j}(\bar{z}_1,...,z_N) \\f^{N...1,\{\chi_i\}}(\bar{z}_1,..,z_j,..,z_N).\label{diffeq1}\end{align}
Note that here $j$ is considered to be a right moving particle without loss of generality. Here the transport operator $Z_{j}(z_1,...,\bar{z}_N)$ transports the particle $j$ around the system once and takes the following form  

  \begin{align}\nonumber
      Z_j(z_1,...,\bar{z}_N)=S'^{jj+1}(z_j,z_{j+1}+L)...S'^{jN}(z_j,z_N+L)\\\nonumber S^{j1}(z_j,\bar{z}_1)... S^{jN_L}(z_j,\bar{z}_{N_L}) S'^{jN_L+1}(z_j,z_{N_L+1})\\...S'^{jj-1}(z_j,z_{j-1}).\label{diffeq2}
  \end{align}  
The transport operators satisfy the following relations
\begin{align}\nonumber
  Z_j(\bar{z}_1,...,z_k-L,...,z_N) Z_k(\bar{z}_1,...,z_N)=\\Z_{k}(\bar{z}_1,...,z_j-L,...,z_N)Z_j(\bar{z}_1,...,z_N). \label{transportop}
\end{align}
In the above equation, we have chosen $j$ and $k$ to be right moving particles. In the case of left moving particles, we simply need to apply the transformation $z_{j/k}\rightarrow\bar{z}_{j/k}$ in the above equation. The constraint equations (\ref{diffeq1}) are matrix difference equations, which need to be solved to obtain the amplitude $f^{N...1,\{\chi_j\}}_{\sigma1...\sigma_N}(\bar{z}_1,...,z_N)$. Once this amplitude is obtained, as mentioned above, one can use the relations (\ref{rel1N}) and (\ref{rel2N}) to obtain the rest of the amplitudes, and thus the explicit form of the $N$ particle wavefunction (\ref{npwf2}). 

\subsection{The case of constant interaction strength and the regular Bethe ansatz technique}
In the previous subsections we have constructed the ansatz wavefunction for time-dependent interaction strength $g(t)$. We have seen that all the amplitudes in the N-particle wavefunction can be expressed in terms of one amplitude of our choosing. By applying periodic boundary conditions, we obtained constraint equations on this amplitude, which take the form of matrix difference equations (\ref{diffeq1}). In this subsection we shall show that in the special case where the interaction strength is constant, this procedure reduces to the standard Bethe ansatz technique, and thereby demonstrating that the above described procedure is a generalization of the regular Bethe ansatz technique.

In the case of constant interaction strength $g(t)\rightarrow g$, as mentioned above (\ref{1conststrength}), the amplitudes in the wavefunction can be expressed in terms of simple exponential functions. For the N-particle wavefunction (\ref{npwf2}), we have

\begin{align} f^{Q,\{\chi_i\}}_{\sigma_1...\sigma_N}(z_1,..,\bar{z}_N)=\prod_{j=N_{L}+1}^{N} e^{ik_jz_j}\prod_{l=1}^{N_L}e^{-ik_jz_j}A^{Q,\{\chi_i\}}_{\sigma_1...\sigma_N},
\end{align}
where $k_j$ are the momenta, and $A^{Q,\{\chi_i\}}_{\sigma_1...\sigma_N}$ are amplitudes corresponding to the ordering of the particles labeled by $Q$, and they do not depend on time $t$ and positions of the particles $x_j$. With this, the N-particle Schrodinger equation turns into time-independent Schrodinger equation which is given by

\begin{align}\nonumber
 &\big(-E-i\hspace{-2.5mm}\sum_{j=N_L+1}^{N}\hspace{-2.5mm}\partial_{x_j}\hspace{-0.5mm}+i\hspace{-0.5mm}\sum_{k=1}^{N_L}\partial_{x_k}\big)\mathcal{A}F^{N...1,\{\chi_i\}}_{\sigma_1...\sigma_N}(x_1,...,x_N)+g\times\\&\sum_{\substack{j=N_L+1\\k=1}}^{\substack{j=N\\k=N_L}}\hspace{-3mm}\delta(x_j-x_k)\vec{\sigma}_{\sigma_j\sigma'_j}\hspace{-1mm}\cdot\hspace{-0.5mm}\vec{\sigma}_{\sigma_k\sigma'_k}\mathcal{A}F^{N...1,\{\chi_i\}}_{\sigma_1..\sigma'_j\sigma'_k..\sigma_N}\hspace{-0.4mm}(x_1,...,x_N)\hspace{-0.7mm}=\hspace{-0.5mm}0,\label{senpti}\end{align}
where the energy 
\be E=\sum_{j=N_L+1}^{N}\hspace{-2.5mm}k_j-\sum_{l=1}^{N_L}k_l. \label{eigen}\ee
The amplitudes which differ by the ordering of the particles with opposite chiralities are related by the S-matrix
\begin{align}
A^{...kj...,\{\chi_i\}}=S^{jk}A^{...jk...,\{\chi_i\}}\label{rel1Nc}
\end{align}
where $\chi_j=+,\chi_k=-$ and $``..."$ in the first superscript on both side of the above equation corresponds to any specific ordering of the rest of the particles. The S-matrix is a constant which only depends on the interaction strength $g$ and is given by 
\begin{align}&\nonumber S^{12}_{ac,bd}=e^{i\phi}\frac{icI^{12}_{ac,bd}+P^{12}_{ac,bd}}{ic+1},\\c=\frac{1}{2g}&\left(1-\frac{3g^2}{4}\right),\;\; e^{i\phi}=\frac{2ig-1+\frac{3g^2}{4}}{ig-\left(1+\frac{3g^2}{4}\right)}.\label{smatlrconst}\end{align}

The S-matrices which relate the amplitudes which correspond to different ordering of the particles with the same chirality are also related by an S-matrix which takes the simple form of the permutation operator
\begin{align}
A^{...kj...,\{\chi_i\}}=P^{jk}A^{...jk...,\{\chi_i\}},\label{rel3Nc}
\end{align}
where $\chi_{j,k}=+$ or $\chi_{j,k}=-$. Just as before, $``..."$ in the first superscript on both side of the above two equations corresponds to any specific ordering of the rest of the particles. The matrix difference equation (\ref{diffeq1}), which is a constraint equation on the amplitudes, now takes the form of an eigenvalue equation \cite{AndreiLowenstein79}

\begin{align}
e^{ik_jL} A^{N...1,\{\chi_i\}}=Z_j A^{N...1,\{\chi_i\}},\label{eigenconst}\end{align}
where the transport operator $Z_j$ takes the form of the transfer matrix
\begin{align}
  Z_j=  P^{jj+1}\dots P^{jN_R} S^{jN_R+1}\dots S^{jN}P^{j1}\dots P^{jj-1}.\label{transportopconst}\end{align}
The relations (\ref{transportop}) turn into simple commutation relations 
\be [Z_i,Z_j]=0,
\ee
which are of fundamental importance in the regular Bethe ansatz method, as they are necessary conditions for the system to be integrable. To solve for the amplitude $A^{N...1,\{\chi_i\}}_{\sigma_1...\sigma_N}$, one diagonalizes the transfer matrix $Z_j$. This can be achieved by the standard algebraic Bethe ansatz technique. One obtains the Bethe equations, whose solutions provide the eigenstates and the corresponding eigenvalues (\ref{eigen}).

\section{Quantum Knizhnik-Zamolodchikov 
equations and the exact wavefunction} 
\label{sec:wavefunction}
In the previous section we have constructed a solution to the N-particle Schrodinger equation (\ref{senp}). The wavefunction consists of amplitudes which correspond to different orderings of the particles with respect to each other. These amplitudes are related to each other through the S-matrices (\ref{rel1N}), (\ref{rel2N}) and (\ref{rel3N}) such that all the amplitudes in the wavefunction can be expressed in terms one amplitude of our choosing. By applying periodic boundary conditions (\ref{pbc}), we obtain constraint equations on this amplitude (\ref{diffeq1}), which take the form of matrix difference equations. In this section we solve these matrix difference equations and obtain the exact solution of the amplitude, and thereby obtain the explicit form of the N-particle wavefunction.  

The matrix difference equation (\ref{diffeq1}) contains S-matrices that relate amplitudes corresponding to different ordering of particles with opposite chiralities and also the particles with the same chiralities. Notice that the S-matrices between particles of opposite chiralities (\ref{smatlr}) have a phase part. In order to solve the equations (\ref{diffeq1}), we need to separate this phase part from the matrix part which acts in the spin spaces of the particles. To achieve this, we apply the following transformation on the amplitudes of the N-particle wavefunction (\ref{npwf2})
\begin{align}
f^{Q,\{\chi_i\}}_{\sigma_1...\sigma_N}(\bar{z}_1,..,z_N)=A^{Q,\{\chi_i\}}_{\sigma_1...\sigma_N}(\bar{z}_1,..,z_N)\hspace{-3mm}\prod_{j=N_{L}+1}^{N}\prod_{l=1}^{N_L}h(\bar{z}_l-z_j).\label{separation}\end{align}

Recall that the interaction strength $g(t)$ should satisfy the constraints (\ref{intconst}) for the solution to be consistent, where the function $f(x)$ is a linear function. Using this, we can express the S-matrices (\ref{smatlr}), (\ref{smatll}) in terms of the XXX R-matrix $R^{ij}(\lambda)$, where
\begin{align}\nonumber
R^{ij}(\lambda)&=\frac{i\lambda I^{ij}+(1/\alpha) P^{ij}}{i\lambda+1/\alpha},\\\nonumber
S^{ij}(z_i,\bar{z}_j)&=e^{i\phi(\bar{z}_j-z_i)}R^{ij}(\bar{z}_j-z_i+\beta/\alpha),\\
S'^{kl}(z_k,z_l)&=R^{kl}(z_k-z_l).\label{relsmatrmat}
\end{align}
Here just as before $I^{ij}$ is the identity matrix and $P^{ij}$ is the permutation operator which acts in the spin spaces of particles $i$ and $j$. Using (\ref{separation}) and (\ref{relsmatrmat}) in the matrix difference equations (\ref{diffeq1}), we see that they can be separated into two sets of equations. The first one concerns only the phase part and takes the form of the analytic difference equation
\begin{align} h(\bar{z}_m-z_j+L)=e^{i\phi(\bar{z}_m-z_j)}h(\bar{z}_m-z_j),\;\; m=1,...,N_L.\label{analyticdiff}
\end{align}
These equations have been well studied in the literature for different classes of functions $\phi(x)$ \cite{ruijisenars}. The second set of equations concern the spin part, which take the form of \textit{quantum Knizhnik-Zamolodchikov equations}. They take the following form

\begin{align}\nonumber
A^{N...1,\{\chi_i\}}(\bar{z}_1,..,z_j-L,..,z_N)=Z'_j(\bar{z}_1,....,z_N) \\A^{N...1,\{\chi_i\}}(\bar{z}_1,..,z_j,..,z_N),\label{qkz}\end{align}
where the transport operator $Z'_j(\bar{z}_1,....,z_N)$ is given by
\begin{align}\nonumber
   &Z'_j(\bar{z}_1,....,z_N)=R^{jj+1}(z_{j+1}+L-z_j)...\\&\nonumber R^{jN}(z_{N}+L-z_j)R^{j1}(\bar{z}_{1}-z_j)...R^{jN_L}(\bar{z}_{N_L}-z_j)\\&R^{jN_L+1}(z_{N_L+1}-z_j)...R^{jj-1}(z_{j-1}-z_j).   \label{qkzop}
   \end{align}
In the context of qKZ equations, $L$ in the above equation is called the \textit{step}. The qKZ equations first appeared in \cite{Smirnov_1986} as the fundamental equations for form factors in the sine-Gordon model, and were later derived from representation theory of quantum affine algebras \cite{Frenkel}. They have been well studied in the literature \cite{rishetikhin1,rishetikhin2,varchenko,matsuo} and the off-shell Bethe ansatz method to solve them has been developed in \cite{Babujian_1997,rishetikhin1}. The solution to these equations provides us with the explicit form of the amplitude $A^{N...1,\{\chi_i\}}(\bar{z}_1, \dots, z_N)$, which can then be used to obtain the rest of the amplitudes and hence also the explicit form of the complete wavefunction (\ref{npwf1}). 

Below, we describe the solution to the qKZ equations (\ref{qkz}), that is obtained following \cite{Babujian_1997}. As mentioned in the beginning, the system conserves the total number of left and right moving fermions separately. We have used these conserved quantities to construct the wavefunction (\ref{npwf1}). In addition to this, our system has the global $SU(2)$ symmetry, and the system conserves the total $z$-component of the spin. This allows us to construct a state with a specific value of $S^z$. Consider a state where the spins of all the particles are pointing in the positive $z$-direction. Let us denote this state by $\ket{\Omega}$ \footnote{This state is referred to as the reference state in the standard Bethe ansatz terminology.}
\be\ket{\Omega}=\ket{\uparrow}_1\otimes...\otimes\ket{\uparrow}_N.
\ee
This state is trivially an eigenstate of the Hamiltonian (\ref{Hamiltonian}) and uninteresting. Now consider a state with $M$ number of spins pointing in the negative $z$-direction and $N-M$ number of spins pointing in the positive $z$-direction . The total $z$-component of the spin corresponding to such as state is
\be S^z=\frac{N}{2}-M.
\ee
There exists an operator $B_0(u_{\alpha})$,  which when acting on the state $\ket{\Omega}$, flips one spin. Here $u_{\alpha}$ is the `rapidity' associated with the spin flip \footnote{The action of one $B_0(u)$ operator on the state $\ket{\Omega}$ produces the state 
\begin{equation}
B_0(u)\, \ket{\Omega}\hspace{-1mm} = \hspace{-1mm}\left(\sum_{i=1}^{N_R} \eta(w_i - u)
\sigma_i^-
+\sum_{i=1}^{N_L} \eta(\bar{w}_i - u)\sigma_i^-\right) \hspace{-0.5mm} \ket{\Omega}
\end{equation}}. A general state with $M$ flipped spins is then constructed by acting on the state $\ket{\Omega}$ by $M$ number of 
$B_0(u_{\alpha})$ operators, where $u_{\alpha}, \alpha=1,...M$ are all distinct. In the case of constant interaction strength, as mentioned before, one constructs eigenstates of the transfer matrix (\ref{eigenconst}), (\ref{transportopconst}), in which case, the set of rapidities $u_{\alpha}, \alpha=1,...,M$ are called the Bethe roots which are solutions to certain constraint equations called the Bethe equations. In the current case, instead of the the eigenstates, we need to construct solutions to the matrix difference equations which take the form of the qKZ equations (\ref{qkz}), (\ref{qkzop}). In this solution, unlike the eigenstates of the transfer matrix, the rapidities are not constrained, but are rather summed over. Following \cite{Babujian_1997}, we obtain
\begin{align}\nonumber
&A^{N...1,\{\chi_i\}}_{\sigma_1...\sigma_N}(\bar{w}_1, \dots, w_N)
= \sum_{u_{\alpha}} \prod_{\alpha=1}^M B_0(u_{\alpha}) \\\nonumber&\hspace{5mm}\times\prod_{i=N_{L}+1}^{N}\prod_{j=1}^{N_L} \prod_{\beta=1}^M \frac{\Gamma(w_i - u_{\beta})}{\Gamma(w_i - u_{\beta} - i\eta)}\frac{\Gamma(\bar{w}_j - u_{\beta})}{\Gamma(\bar{w}_j - u_{\beta} - i\eta)} \\&\hspace{5mm}\times\prod_{1 \le i < j \le M} \frac{(u_i - u_j)\, \Gamma(u_i - u_j - i\eta)}{\Gamma(u_i - u_j + i\eta + 1)} \ket{\Omega},
\label{ansatzstatefinalform}
\end{align}
where the summation is over the integers $l_{\alpha}$, while the parameters $\widetilde{u}_{\alpha}$, $\alpha = 1, \dots, M$, are arbitrary constants
\be 
u_{\alpha} = \widetilde{u}_{\alpha} - l_{\alpha}, \quad l_{\alpha} \in \mathbb{Z}.
\label{summation}
\ee
This infinite sum is called a `Jackson type integral'. In the above expression (\ref{ansatzstatefinalform}), $\Gamma(x)$ is the usual Gamma function and $\eta=1/(\alpha L)$. The parameters $w_i$, $\bar{w}_i$ and $\eta$ are related to $z_i$, $\bar{z}_i$, $\alpha$ and $\beta$ through the following relations 
\begin{align}\nonumber
   w_i&=\frac{z_i}{L}, \;\;\hspace{10mm}i=N_{L+1},...,N;\\\bar{w}_i&=\frac{\bar{z}_i}{L}+\frac{\beta}{\alpha L}, \label{relGNqKZ}\;\;i=1,...,N_L. 
\end{align}
Note that, in addition to the qKZ equations (\ref{qkz}), we need to solve the analytic difference equation (\ref{analyticdiff}) to obtain the function $h(x)$. The solution is complicated, but it simplifies in the limit $\alpha,\beta\gg1$, where the interaction strength $g(t)$ (\ref{intstrength}) takes the form
\be g(t)=\frac{1}{4(\alpha t+\beta/2)}.
\ee
Here we have chosen the positive sign in (\ref{intstrength}). We note that in the limit of large times, the above expression corresponds to small values of the interaction strength. This precisely corresponds to the universal regime mentioned previously, where different regularizations yield the same form of the interaction strength.

The function $h(x)$ is given by
\begin{align}
    h(x)=e^{2i\pi nx/L}\frac{\Gamma((x+\beta/\alpha-i/\alpha)/L)}{\Gamma((x+\beta/\alpha-i/2\alpha)/L)},
\end{align}
where $n$ is an integer. Instead of the exponential function in the above expression, one may choose any function whose period is commensurate with $L$.  Having obtained the explicit form of the amplitude $f^{N...1,\{\chi_i\}}_{\sigma_1...\sigma_N}(\bar{w}_1, \dots, w_N)$, the rest of the amplitudes and hence the explicit form of the wavefunction (\ref{npwf1}) can be obtained from it by the action of S-matrices (\ref{rel1N}), (\ref{rel2N}) and (\ref{rel3N}). 

\section{Discussion}
\label{sec:discussion}
In this work we have considered the time-dependent $SU(2)$ Gross-Neveu model. In this quantum field theory, spin $1/2$ fermions interact with each other through spin exchange interaction strength that varies in time.  Using the recently formulated framework \cite{PasnooriKondo}, which generalizes the standard Bethe ansatz technique, we have constructed an exact solution to the time-dependent Schrodinger equation. We considered the system with periodic boundary conditions. This results in the conservation of the number of left and right moving fermions separately, thus allowing us to use them to label the wavefunction. The wavefunction consists of several amplitudes, where each amplitude corresponds to a certain ordering of particles with respect to each other. Any amplitude in the wavefunction can be related to one amplitude of our choosing through the action of particle-particle S-matrices. The amplitudes and the S-matrices contain a phase part and a spin part. The consistency of the wavefunction requires that these S-matrices satisfy Yang-Baxter equation, which is a necessary condition for the integrability of the system. This imposes constraints on the time-dependent interaction strength. In the regular Bethe ansatz approach, which is applicable to Hamiltonians with constant interaction strengths, the amplitudes and the S-matrices are constants. The phase part of the amplitudes take the form of simple exponential functions. In contrast, in our case of time-dependent interaction strength, these amplitudes are vector valued functions which depend on the time depend interaction strength and also on the positions of all the particles. The chosen amplitude is then determined by applying periodic boundary conditions, which gives rise to matrix difference equations which constrain the amplitude. We showed that these matrix difference equations reduce to a set of analytic difference equations which govern the phase part, and quantum Knizhnik-Zamolodchikov (qKZ) equations which govern the spin part. 

In the case of the constant interaction strength, these matrix difference equations reduce to an eigenvalue equation involving a transfer matrix, which is then diagonalized using the standard algebraic Bethe ansatz technique. Hence, one can consider the regular Bethe ansatz method as a special case of the general Bethe ansatz method used in this work. In our case of time-dependent interaction strength, as mentioned above, the general Bethe ansatz method gives rise to qKZ equations and analytic difference equations. The qKZ equations were solved using the off-shell algebraic Bethe ansatz technique. Using the solution to the qKZ equations, along with the solution to the analytic difference equations, we obtained the explicit form of the amplitude. The rest of the amplitudes can then be straightforwardly determined by the action of the particle-particle S-matrices on this amplitude, and thereby one can obtain the explicit form of the complete many-body wavefunction.

In addition to the $SU(2)$ symmetric case considered in this work, one may consider the case where the $SU(2)$ symmetry is broken down to $U(1)$ symmetry. In which case one obtains the $U(1)$ Thirring model with time-dependent interaction strength. This case is expected to be more interesting since it contains two coupling strengths which vary in time. In this case, instead of the qKZ equations corresponding to the XXX-R matrix that we obtained in this work, one obtains the qKZ equations corresponding to the XXZ R-matrix \cite{rishetikhin2,pasnoori1}. In addition, one may consider different boundary conditions as opposed to simple periodic boundary conditions considered in this work. Under these boundary conditions, in addition to the bulk interaction strengths, there exist boundary coupling strengths which can vary in time. In the simple case where the bulk interaction strengths and the boundary coupling strengths are constant, the models described above are shown to exhibit symmetry protected topological (SPT) phases \cite{PAA1,PAA2,pasnoori2025duality,pasnoori2025interplay}. Hence, the case where the bulk interaction strengths and the boundary coupling strengths vary in time is naturally very interesting, as they may give rise to new type of dynamically generated SPT phases, which is the focus of our future work \cite{pasnoori2}. We finally emphasize that time-dependent quantum field theories of the type studied here can be naturally realized in superconducting quantum circuits constructed from networks of Josephson junctions. Away from the SU(2)-symmetric point, the Gross-Neveu model reduces to the U(1) Thirring model, whose spin sector is described by the sine-Gordon model. Importantly, both the sine-Gordon model and the U(1) Thirring model with static coupling strengths have already been demonstrated to be implementable in superconducting circuit architectures \cite{pasnoori2025circuitSG, SGroy, MMpasnooricircuit}. In these realizations, the effective coupling constants of the field theories are directly related to circuit parameters such as inductances and Josephson energies, which can be dynamically tuned in time. This tunability therefore provides a concrete experimental realization of time-dependent quantum field theories within superconducting circuit platforms.

\section*{Acknowledgments}
We acknowledge discussions with Natan Andrei, Patrick Azaria, Paul Fendley and David Huse. I specially thank Patrick Azaria for carefully reviewing the manuscript.

\bibliography{refpaper}
\begin{appendix}
 \begin{widetext}
\numberwithin{equation}{section}

\section{Solutions to the quantum Knizhnik-Zamolodchikov equations}
\label{appendix}

In the main text, we have constructed an ansatz wavefunction for the N-particle Schrodinger equation. We have seen that the ansatz wavefunction consists of amplitudes which correspond to different orderings of the particles with respect to each other and are related by S-matrices. All the amplitudes in the wavefunction can be expressed in terms of any one amplitude of our choosing.  We have shown that by applying periodic boundary conditions to the N-particle wavefunction results in a constraint equation on this amplitude, which takes the form of a set of matrix difference equations.  We have further shown that these matrix difference equations can be reduced to a set of analytic difference equations and quantum Knizhnik-Zamolodchikov equations. The general form of qKZ equations is given by

\be
A^{N...1,\{\chi_i\}}(\bar{w}_1\dots, w_j + \kappa,\dots, w_N) 
= Z'_j(\bar{w}_1,\dots,w_N)\, 
A^{N...1,\{\chi_i\}}(\bar{w}_1,\dots,w_N),
\label{qKZap}
\ee
where the transport operator $Z'_j(\bar{w}_1,\dots,w_N)$ is given by
\begin{align}\nonumber
Z'_j(\bar{w}_1,\dots,w_N) &= 
R^{j+1j}(w_{j+1} - w_{j}-\kappa)... R^{Nj}(w_N-w_{j}-\kappa) R^{1j}(\bar{w}_{1}-w_j)\cdots \\&R^{N_Lj}( \bar{w}_{N_L}-w_j) R^{N_{L}+1j}(w_{N_L+1}-w_j) \cdots R^{j-1j}(w_{j-1}-w_j),
\label{transportopap}
\end{align}
where $R^{jk}(x)$ is the XXX R-matrix, which is given by
\begin{equation}
R^{jk}(x) = \lambda(x)\, I^{jk} + \eta(x)\, P^{jk}, \qquad
\lambda(x) = \frac{ix}{ix + \eta}, \qquad
\eta(x) = \frac{\eta}{ix + \eta},
\label{rmatap}
\end{equation}
where $\eta$ is the crossing parameter, $I^{jk}$ is the identity operator, and $P^{jk}$ is the permutation operator acting on the tensor product of spin spaces labeled by $j$ and $k$. $\kappa$ is the step, which is related to the time-dependent Gross-Neveu model by $\kappa=-1$. Here, we shall work with the general form of the qKZ equations and construct an explicit solution to these equations using the off-shell Bethe ansatz method. The parameters $w_j, j=N_{L}+1...N$, $\bar{w}_k, k=1,...,N_L$ and the crossing parameter $\eta$ are related to the parameters of the time-dependent Gross-Neveu model through the relations presented in the main text (Please see Eq. (\ref{relGNqKZ}) and the text above it). We note that $w_j$ and $\bar{w}_k$ are related to $z_j$ and $\bar{z}_k$ such that the amplitudes 
$A^{Q,\{\chi_i\}}(\bar{w}_1\dots, w_j,\dots, w_N)$ also satisfy the periodicity condition (\ref{nppbc}).  

The off-shell Bethe ansatz method to solve the qKZ equations was originally developed by Babujian \cite{Babujian_1997}. Here, we reproduce the essential steps for completeness. The procedure for solving the qKZ equations essentially consists of two steps:

\vspace{1mm}

\noindent 1) Introduce shifted monodromy matrix and Q-monodromy matrix, and obtain their respective operator algebras.

\vspace{1mm}

\noindent 2)Choose an ansatz for the amplitude and use the operator algebra obtained in the first step to obtain the solution to the qKZ equations. 

\vspace{1mm}

\noindent These steps are described in detail in separate subsections below.

\subsection{Operator algebra}
\label{sec_monodromy}

In this section we first introduce the shifted monodromy matrix, which is related to the standard monodromy matrix used in the standard algebraic Bethe ansatz procedure, and obtain the operator algebra asscoaited with it. We then introduce a new type of monodromy matrix first introduced by Babujian in \cite{Babujian_1997}, which we call Q-monodromy matrix. We discuss its properties and obtain the operator algebra associated with it. 

\subsubsection{Monodromy matrix}

Let us consider a tensor product space \[
 V_1 \otimes \cdots \otimes V_N \otimes V_a,
\]
where $V_1, \dots, V_N$ correspond to the $N$ particles and $V_a$ is the auxiliary space. Since we are working with spin-$\frac{1}{2}$ particles, each space $V_\alpha \cong \mathbb{C}^2$ for $\alpha = a, 1, \dots, N$. One defines a `shifted' monodromy matrix which acts in the tensor product space mentioned above. It takes the following form
\begin{equation}
\label{monodromyj}
T^j_{N\dots 1\, a}(w) = R_{1a}(\bar{w}_1 - w)\cdots R_{ja}(\bar{w}_j-\kappa-w)\cdots R_{N_La}(\bar{w}_{N_L} - w)R_{N_{L+1}a}(w_{N_L+1}-w)...R_{Na}(w_N-w).
\end{equation}
Here, $w\in \mathcal{C}$ is called the spectral parameter and $R_{ia}(w_i-w)$ is the R-matrix (\ref{rmatap}) which acts in the spin spaces $i$ and $a$.  $\{w_i\} \equiv (\bar{w}_1, \dots,\bar{w}_{N_L},w_{N_L+1} \dots, w_N)$ represents the set of parameters associated with the transport operator $Z'_j(\bar{w}_1,...,w_N)$ (\ref{transportopap}). The superscript in $T^j_{N\dots 1\, a}(w)$ denotes that the parameter $w_j$ in the set of parameters $\{w_i\}$ is shifted by one step unit $\kappa$. Note that $j$ takes values $1...N$ in the above expression. For $j=0$, we have
\begin{equation}
\label{monodromy}
T^0_{N\dots 1\, a}(w) = R_{1a}(\bar{w}_1 - w)\cdots R_{N_La}(\bar{w}_{N_L} - w)R_{N_{L+1}a}(w_{N_L+1}-w)...R_{Na}(w_N-w).
\end{equation}
One can recognize that this is the standard monodromy matrix of the algebraic Bethe ansatz. Let us now recollect some important properties of the XXX $R$-matrix, that we shall use in the following. 

\begin{align}
& \text{Initial condition:} \; R_{ab}(0) = P_{ab},\\
   &\text{Unitarity:} \; R_{ab}(x) R_{ab}(-x) = \mathbb{1},\\
 &\text{Crossing symmetry:}\; 
  R_{ab}(x) = -\sigma^y_a R^{t_a}_{ab}(-x + i\eta)\, \sigma^y_a \left( \frac{x}{-x + i\eta} \right),
  \end{align}

where $t_a$ denotes transposition in space $a$ and $\sigma^y_a$ is the Pauli matrix acting in that space.

The $R$-matrices satisfy the Yang--Baxter equation:
\begin{equation}
R^{ij}(w_i - w_j)\, R^{ik}(w_i - w_k)\, R^{jk}(w_j - w_k)
= R^{jk}(w_j - w_k)\, R^{ik}(w_i - w_k)\, R^{ij}(w_i - w_j).
\label{YB1ap}
\end{equation}
 Using these relations, one can show that the monodromy matrices (\ref{monodromy}) with different spectral parameters also satisfy the Yang--Baxter equation:
\begin{equation}
T^0_{N \dots 1\, a}( u)\, T^0_{N \dots 1\, b}( v)\, R_{ab}(u - v) 
= R_{ab}(u - v)\, T^0_{N \dots 1\, b}( v)\, T^0_{N \dots 1\, a}( u),
\label{YBMonodromy}
\end{equation}
where $a$ and $b$ are auxiliary spaces. Using the monodromy matrix, one defines the transfer matrix
\begin{align}
    tr_a T^0_{N \dots 1\, a}( u)=t^0_{N \dots 1}(u)\end{align}
Using the relations (\ref{YBMonodromy}), one can show that the transfer matrices with different spectral parameters commute
\begin{align}
    \big[t^0_{N \dots 1}(u),t^0_{N \dots 1}(v) \big]=0.  \label{commutransfer} \end{align}
In the regular Bethe ansatz, which is applicable to systems whose parameters are independent of time, the property (\ref{commutransfer}) ensures that the system is integrable. In our case where the parameters of the system are dependent on time, in addition to (\ref{commutransfer}), the system satisfies additional properties, to be discussed below, which are responsible for the integrability of our model. 

 Let the shifted monodromy matrix take the following form in the auxiliary space:
\begin{equation}
T^j_{N \dots 1\, a}(u) =
\begin{pmatrix}
A_j( u) & B_j( u) \\
C_j( u) & D_j( u)
\end{pmatrix},
\label{Monoexpl}
\end{equation}
where the operators $X_j( u)$, with $X = A, B, C, D$, act in the tensor product space $V_1 \otimes \cdots \otimes V_N$ corresponding to the particles ($1,\dots,N$). For $j=0$ in (\ref{Monoexpl}), one obtains the explicit form of the monodromy matrix. Using this in the relation (\ref{YBMonodromy}), one obtains the following commutation relations associated with the operators $A_0( u), B_0( u), C_0(u)$ and $D_0( u)$. 
\begin{align}\label{commrel1ap}
A_0(u)\, A_0(v) &= A_0(v)\, A_0(u), \\
B_0(u)\, B_0(v) &= B_0(v)\, B_0(u), \\
C_0(u)\, C_0(v) &= C_0(v)\, C_0(u), \\
D_0(u)\, D_0(v) &= D_0(v)\, D_0(u),
\end{align}
\begin{align}
A_0(\mu)B_0(\nu)&=\frac{1}{\lambda(\mu-\nu)}B_0(\nu)A_0(\mu)-\frac{\eta(\mu-\nu)}{\lambda(\mu-\nu)}B_0(\mu)A_0(\nu)\\
D_0(\mu)B_0(\nu)&=\frac{1}{\lambda(\nu-\mu)}B_0(\nu)D_0(\mu)-\frac{\eta(\nu-\mu)}{\lambda(\nu-\mu)}B_0(\mu)D_0(\nu)
\label{commrelMonoap}
\end{align}
where the scalar functions $\lambda(x)$ and $\eta(x)$ are defined in (\ref{rmatap}). The relations (\ref{commrelMonoap}) obtained above are the standard commutation relations associated with the XXX R-matrix that one encounters in the algebraic Bethe ansatz \cite{SklyaninQISM}.

\subsubsection{Q-monodromy matrix}
In the previous subsection, we obtained the commutation relations associated with the regular monodromy matrix. Here, following \cite{Babujian_1997}, we shall introduce a new monodromy matrix which we shall call `Q-monodromy matrix'. This acts in the same tensor product space associated with the spin spaces of the particles and the auxiliary space mentioned in the previous subsection. The Q-monodromy matrix takes the following form

\begin{equation}
T^{Q_j}_{N\dots 1\,a}(w) = R_{1a}(\bar{w}_1 - w) \dots R_{N_L\,a}(\bar{w}_{N_L} - w)\dots R_{j\,a}(w_j - w)\, R_{j+1\,a}(w_{j+1} - w - \kappa) \dots R_{N\,a}(w_N - w -\kappa),
\label{Qmonodromy}
\end{equation}
 The superscript $Q_j$ on the left side of the above equation denotes that the parameters $w_i$ in the R-matrices acting in the spin spaces of particles $j+1,...,N$ are shifted by one step unit. The transport operator (\ref{transportopap}) is related to the Q-monodromy matrix through the following relation

\begin{equation}
Z'_j = \mathrm{tr}_a\, T^{Q_j}_{N\dots 1\,a}(w_j).
\label{relzt}
\end{equation}

The Q-monodromy matrix (\ref{Qmonodromy}) and the regular monodromy matrix (\ref{monodromy}) are in general different, but for $j=N$ in (\ref{Qmonodromy}), one can see that the Q-monodromy matrix exactly coincides with the standard monodromy matrix

\begin{equation}
T^{Q_N}_{N\dots 1\,a}( w) = T^0_{N\dots 1\,a}( w).
\label{equivtmats}
\end{equation}

The Q-monodromy matrix and the regular monodromy matrix satisfy the following new type of Yang-Baxter equation

\begin{align}
&T^{Q_j}_{N\dots 1\,a}( w_j)\, T^0_{N\dots 1\,b}( w)\, R_{ab}(w_j+\kappa - w) = R_{ab}(w_j - w)\, T^j_{N\dots 1\,b}( w)\, T^{Q_j}_{N\dots 1\,a}( w_j).
\label{newYB}
\end{align} 
Recollect that $T^j_{N\dots 1\,b}( w)$ is the shifted monomdromy matrix (\ref{monodromyj}). Let the Q-monodromy matrix take the following form in the auxiliary space:
\begin{equation}
T^{Q_j}_{N\dots 1\,a}( w) =
\begin{pmatrix}
A^{Q_j}( w) & B^{Q_j}(w) \\
C^{w_j}(w) & D^{Q_j}(w)
\end{pmatrix}.
\label{Qmonoexpl}
\end{equation}
Using the explicit form of the Q-monodromy matrix (\ref{Qmonoexpl}) and the shifted monodromy matrix (\ref{Monoexpl}), in the new type of Yang-Baxter relation(\ref{newYB}), we obtain the following commutation relations among the operators $X^{Q_j}(w)$ and $X_j(w)$ with $X = A, B, C, D$. 
\begin{align}\label{commrel3ap}
A^{Q_j}(w_j) A_0(w) &= A_j(w) A^{Q_j}(w_j), \\
C^{Q_j}(w_j) C_0(w) &= C_j(w) C^{Q_j}(w_j), \\
B^{Q_j}(w_j) B_0(w) &= B_j(w) B^{Q_j}(w_j), \\
D^{Q_j}(w_j) D_0(w) &= D_j(w) D^{Q_j}(w_j), \\
A^{Q_j}(w_j) B_0(w) &= \frac{1}{\lambda(w_j - w + \kappa)} B^{Q_j}(w_j) A_0(w)
+ \frac{1}{\eta(w_j - w + \kappa)} A_j(w) B^{Q_j}(w_j), \\
A^{Q_j}(w_j) B_0(w) &= \frac{1}{\lambda(w_j - w + \kappa)} B_j(w) A^{Q_j}(w_j)
- \frac{\eta(w_j - w + \kappa)}{\lambda(w_j - w + \kappa)} B^{Q_j}(w_j) A_0(w), \\
C^{Q_j}(w_j) A_0(w) &= \lambda(w_j - w) A_j(w) C^{Q_j}(w_j)
+ \eta(w_j - w) C_j(w) A^{Q_j}(w_j), \\
A^{Q_j}(w_j) C_0(w) &= \lambda(w_j - w) C_j(w) A^{Q_j}(w_j)
+ \eta(w_j - w) A_j(y) C^{Q_j}(w_j), \\
D^{Q_j}(w_j) B_0(w) &= \frac{1}{\lambda(w - w_j)} B_j(w) D^{Q_j}(w_j)
- \frac{\eta(w - w_j)}{\lambda(w - w_j)} B^{Q_j}(w_j) D_0(w).
\label{commrelqMonoap}
\end{align}
Unlike the standard commutation relations associated with the regular monodromy matrix (\ref{commrelMonoap}), these commutation relations which are associated with the new type of Yang-Baxter relation (\ref{newYB}) are new and were originally obtained in \cite{Babujian_1997}.

\subsection{Ansatz for the amplitude}
In the previous section, we have introduced the shifted monodromy and the Q-monodromy matrices, which are abstract operators acting in the tensor product space consisting of the spin spaces of the particles and the auxiliary space. Using the Yang-Baxter (\ref{YBMonodromy}) and the new Yang-Baxter relations (\ref{newYB}), we have derived the commutation relations of the operators associated with these matrices. In this section, we shall use these relations to solve the qKZ equations (\ref{qKZap}). Before venturing into this, let us very briefly discuss the standard Bethe ansatz procedure which is used in the case where the interaction strength is a constant. 

\subsubsection{Standard Bethe ansatz for the case of constant interaction strength}
In the case where the interaction strength is a constant, we have seen in the maintext that the constraint equation on the amplitude takes the form of an eigenvalue equation (\ref{eigenconst}). Solving this equation provides the amplitude $A^{N...1,\{\chi_i\}}_{\sigma_1...\sigma_N}$, which corresponds to a particular eigenstate. To solve this equation, one uses the method of algebraic Bethe ansatz technique. The first step is to relate the transport operator (\ref{transportopconst}) to the monodromy matrix (\ref{monodromy}). In this case of constant interaction strength, the parameters in the monodromy matrix are constants, where $w_i=b, i=N_L+1...N$ and $\bar{w}_i=-b, i=1,...,N_L$. We represent this set of parameters by $\{w_i\}_b$. The monodromy matrix is related to the transport operator (\ref{transportopconst}) through the relation
\begin{align}
    tr_a T^0_{N...1\,a}(b)=Z_1.\label{trmonoconst}
\end{align}
Here $Z_1$ transports the particle 1 across the entire system once, and $b$ is a constant which is related to the constant interaction strength $g$ \cite{PAA1,PAA2}. As mentioned above, one needs to solve the eigenvalue equation (\ref{eigenconst}). This requires one to diagonalize the trace of the regular monodromy matrix (\ref{trmonoconst}). To achieve this, one first considers the reference state $\ket{\Omega}$ in which the spins of all particles point along the positive $z$-direction:
\begin{equation}
\ket{\Omega} = \ket{\uparrow \dots \uparrow}.
\end{equation}
This is an eigenvector of the operators $A_0( u)$ and $D_0(u)$ associated with the monodromy matrix (\ref{Monoexpl}), where
\begin{align}
A_0( u)\, \ket{\Omega} &= \ket{\Omega}, \label{eigenvalueA} \\
D_0( u)\, \ket{\Omega} &= \prod_{i=1}^{N_R}\prod_{j=1}^{N_L} \lambda(w_i - u)\lambda(\bar{w}_j-u)\, \ket{\Omega}. \label{eigenvalueD}
\end{align}
 The operator $C_0(u)$ acts trivially on the reference state:
\begin{equation}
C_0( u)\, \ket{\Omega} = 0,
\end{equation}
while the operator $B_0( u)$ acts nontrivially, flipping one spin:
\begin{equation}
B_0(u)\, \ket{\Omega} = \left(\sum_{i=1}^{N_R} \eta(w_i - u)
\sigma_i^-
+\sum_{i=1}^{N_L} \eta(\bar{w}_i - u)\sigma_i^-\right)  \ket{\Omega}.
\end{equation}

In the standard Bethe ansatz, one constructs an ansatz state with $M$ spins pointing in the downward $z$-direction by acting on $\ket{\Omega}$ with $M$ spin-flipping operators $B_0( u_\alpha)$, where $\alpha = 1, \dots, M$:
\begin{equation}
\ket{\Omega'}_M = \prod_{\alpha=1}^{M} B_0( u_\alpha)\, \ket{\Omega}.
\label{eigenstateap}
\end{equation}
This state has total $z$-component of spin
\begin{equation}
S^z = \frac{N}{2} - M.
\end{equation}
One applies the trace of the monodromy matrix \ref{trmonoconst} to the above state (\ref{eigenstateap})

\begin{align}
    \big(A_0(b)+D_0(b)\big)\ket{\Omega'}_M.\end{align}
Note that the reference state $\ket{\Omega}$ is an eigenstate of the operators $A_0(b)$ and $D_0(b)$. One commutes these operators past the $B_0( u_\alpha)$ operators in the state $\ket{\Omega'}_M$ (\ref{eigenstateap}) and uses the commutation relations associated with the monodromy matrix (\ref{commrelMonoap}). This procedure gives rise to several terms which contain wanted terms which take the form of eigenstates. In addition, they contain unwanted terms which cannot be expressed in terms of eigenstates. Requiring that these unwanted terms vanish gives rise to constraint equations between the parameters $u_\alpha$, which are called \textit{Bethe roots}. They must all be distinct in order for the state $\ket{\Omega'}_M$ to be nonzero. These constraint equation on the Bethe roots are called the Bethe equations. Every solution to these equations provides a set of Bethe roots $\{u_{\alpha}\}$. Each eigenstate of the system corresponds to a particular set of Bethe roots.  For each solution set of Bethe roots, the coefficient of the wanted terms provides the eigenvalue. In the case of time-dependent interaction strength, the procedure is very different, which we shall discuss below.

\subsubsection{The case of time-dependent interaction strength}
In the current case where the interaction strength is not constant, the standard algebraic Bethe ansatz briefly discussed in the previous subsection does not apply. This can be anticipated as the constraint equation on the amplitude is instead a set of matrix difference equations which take the form of quantum Knizhnik-Zamolodchikov (qKZ) equations (\ref{qKZap}). Below we provide the solution to the qKZ equations, which is achieved using the off-shell Bethe ansatz method.

Consider the qKZ equations (\ref{qKZap}). The transport operator (\ref{transportopap}) on the right side of the qKZ equations is related to the trace of the Q-monodromy matrix
\begin{align}
Z'_j = \text{tr}_a\, T^{Q_j}_{N\dots 1\,a}( w_j) = A^{Q_j}(w_j) + D^{Q_j}(w_j).
\label{traceqmono}
\end{align}

We now have to choose an ansatz for the amplitude $A^{N...1,\{\chi_i\}}(\bar{w}_1,\dots,w_N)$. Consider the following form:
\begin{align}
A^{N...1,\{\chi_i\}}(\bar{w}_1,\dots, w_N)=\sum_{u_{\alpha}}\prod_{\alpha=1}^{M}B_0(u_{\alpha})\,v(\{w_i\},\{u_{\alpha}\})\,\ket{\Omega}.
\label{ansatzstate}
\end{align}
The total $z$-component of the spin of this state is $S^z = \frac{N}{2} - M$. 
Here, $v(\{w_i\},\{u_{\alpha}\})$ is yet an unknown function, that we need to determine. The parameters $u_{\alpha}, \alpha=1,...,M$ are denoted by $\{u_{\alpha}\}$. Note that unlike the ansatz state in the regular Bethe ansatz procedure (\ref{eigenstateap}), where the parameters $\{u_{\alpha}\}$ correspond to Bethe roots that satisfy Bethe equations, here in the ansatz (\ref{ansatzstate}) they are summed over:
\be 
u_{\alpha} = \widetilde{u}_{\alpha} - l_{\alpha}, \quad l_{\alpha} \in \mathbb{Z}.
\label{summation}
\ee
The summation is over the integers $l_{\alpha}$, while the parameters $\widetilde{u}_{\alpha}$, $\alpha = 1, \dots, M$, are arbitrary constants.

Since the amplitudes $A^{N...1,\{\chi_i\}}_{\sigma_1...\sigma_N}(\bar{w}_1,...,w_N)$ also satisfy the periodicity condition (\ref{nppbc}), it suffices to consider only the transport operator $Z'_N$. Consider the right side of the qKZ equation (\ref{qKZap}). This corresponds to the action of the transport operator, or equivalently, the trace of the Q-monodromy matrix on the ansatz state (\ref{ansatzstate}). Similar to the regular algebraic Bethe ansatz, one commutes the operators $A^{Q_j}( w_j)$ and $D^{Q_j}( w_j)$ in the trace of the Q-monodromy matrix past the operators $B_0(u_j)\,v(\{w_i\},\{u_{\alpha}\})$ in the ansatz state (\ref{ansatzstate}).  In this process, one uses the commutation relations corresponding to the monodromy matrix (\ref{commrelMonoap}) and Q-monodromy matrix (\ref{commrelqMonoap}) and obtains several terms which can be categorized as `wanted' and `unwanted'. The wanted terms take the form of the left side of the qKZ equations, whereas the unwanted terms do not. For detailed discussion of this procedure, we refer the reader to the original work \cite{Babujian_1997}. Carrying out this procedure, we obtain the following expression:
\begin{align}\nonumber
&\text{tr}_a\, T^{Q_N}_{N\dots 1\, a}(w_N)\, A^{N...1,\{\chi_i\}}(\bar{w}_1, \dots, w_N) = \left(A^{Q_N}(w_N) + D^{Q_N}( w_N)\right) A^{N...1,\{\chi_i\}}(\bar{w}_1, \dots, w_N)\\\nonumber
&= \sum_{u_{\alpha}} \prod_{\alpha=1}^M B_N( u_{\alpha})\, v(\{w_i\}, \{u_{\alpha}\}) \left[ \prod_{\beta=1}^M \frac{1}{\lambda(w_N + \kappa - u_{\beta})} A^{Q_N}( w_N) + \prod_{\beta=1}^M \frac{1}{\lambda(u_{\beta} - w_N)} D^{Q_N}( w_N) \right] \ket{\Omega}\\& + \sum_{u_{\alpha}} \sum_{m=1}^M \left(UW_A^m + UW_D^m\right),
\label{ansatzstateaction}
\end{align}
where the last two terms, $UW_A^m$ and $UW_D^m$, are the unwanted terms, which are given by
\begin{align}
UW_A^m = -\frac{\eta(w_N + \kappa - u_m)}{\lambda(w_N + \kappa - u_m)}\, B^{Q_N}( w_N) \prod_{\substack{\alpha=1 \\ \alpha\ne m}}^M B_0( u_{\alpha}) \prod_{\beta\neq m, \beta=1}^M \frac{1}{\lambda(u_m - u_{\beta})}\, A_0( u_m)\, v(\{w_i\}, \{u_{\alpha}\}) \ket{\Omega},
\end{align}
\begin{align}
UW_D^m = -\frac{\eta(u_m - w_N)}{\lambda(u_m - w_N)}\, B^{Q_N}(w_N) \prod_{\substack{\alpha=1 \\ \alpha\ne m}}^M B_0( u_{\alpha}) \prod_{\beta\neq m, \beta=1}^M \frac{1}{\lambda(u_{\beta} - u_m)}\, D_0(u_m)\, v(\{w_i\}, \{u_{\alpha}\}) \ket{\Omega}.
\end{align}
Using the expression for the action of the operator $D^{Q_j}(u)$ on the reference state $\ket{\Omega}$ (\ref{eigenvalueD}), one can see that the second term in the above expression vanishes  $D^{Q_N}( w_N)\ket{\Omega} = 0$.  The first term has the same structure as the ansatz state itself as it contains only $B_N(u_{\alpha})$ operators, while the last two terms $UW_A^m$ and $UW_D^m$ do not, as they involve operators $B^{Q_N}(w_m)$. Hence in order to obtain a solution to the qKZ equations, these unwanted terms should be eliminated. This can be achieved by appropriately choosing the function $v(\{w_i\}, \{u_{\alpha}\})$ such that the unwanted terms cancel pairwise. Consider the following ansatz:
\be
v(\{w_i\}, \{u_{\alpha}\}) = \prod_{i=1}^{N_R}\prod_{j=1}^{N_L} \prod_{\alpha=1}^M \xi(w_i - u_{\alpha})\xi(\bar{w}_j-u_{\alpha})\!\!\! \prod_{1 \le \beta < \alpha \le M} \tau(u_{\beta} - u_{\alpha}),
\label{defv}
\ee
where the functions $\xi(x)$ and $\tau(x)$ satisfy the following relations
\be
\lambda(x)\, \xi(x) = \xi(x - \kappa), \qquad \frac{\tau(x)}{\lambda(x)} = \frac{\tau(x - \kappa)}{\lambda(\kappa - x)}.
\label{constraintsxitau}
\ee
One can verify that this ansatz gives rise to the following relation between the unwanted terms
\be
UW_A^m(\{w_i\}, \{u_{\alpha}\}_m)\, v(\{w_i\}, \{u_{\alpha}\}_m) = -UW_D^m(\{w_i\}, \{u_{\alpha}\})\, v(\{w_i\}, \{u_{\alpha}\}),
\label{unwantedrel}
\ee
Here, $\{u_{\alpha}\}_m$ denotes the set of parameters $u_{\alpha}, \alpha=1,...,M$ with $u_m$ replaced by $u_m + \kappa$. One can see that using the relation (\ref{unwantedrel}) along with the summation (\ref{summation}) in the expression (\ref{ansatzstateaction}), the unwanted terms cancel each other leaving only the first term. We have
\be
\text{tr}_a\, T^{Q_N}_{N\dots 1\,a}( w_1)\,A^{N...1,\{\chi_i\}}(\bar{w}_1, \dots, w_N) = \sum_{u_{\alpha}} \prod_{\alpha=1}^M B_N( u_{\alpha})\, v(\{w_i\}, \{u_{\alpha}\}) \prod_{\beta=1}^M \frac{1}{\lambda(w_N + \kappa - u_{\beta})} \ket{\Omega}.
\label{solutionqKZ}
\ee
Using the functional relation (\ref{constraintsxitau}), we confirm that the right-hand side of equation~(\ref{solutionqKZ}) matches the left-hand side of the qKZ equation (\ref{qKZap}). Therefore,
\be
Z'_N\,A^{N...1,\{\chi_i\}}(\bar{w}_1, \dots, w_N)= A^{N...1,\{\chi_i\}}(\bar{w}_1, \dots, w_N+\kappa)
\ee
 This establishes that the ansatz (\ref{ansatzstate}) solves the qKZ equation (\ref{qKZap}), provided the functions $\xi(x)$ and $\tau(x)$ satisfy the constraints~(\ref{constraintsxitau}). Solving these functional equations using the definition of $\lambda(x)$ from (\ref{rmatap}), we find the explicit solutions
\be
\xi(x) = \frac{\Gamma(x)}{\Gamma(x - i\eta)}, \qquad \tau(x) = x\, \frac{\Gamma(x - i\eta)}{\Gamma(x + i\eta + 1)}.
\label{solutionxitau}
\ee
Substituting these expressions in (\ref{defv}), we obtain the final form of the amplitude:
\begin{align}\nonumber
A^{N...1,\{\chi_i\}}(\bar{w}_1, \dots, w_N) = \sum_{u_{\alpha}} \prod_{\alpha=1}^M B_0( u_{\alpha}) \prod_{i=1}^{N_R}\prod_{j=1}^{N_L} \prod_{\beta=1}^M \frac{\Gamma(w_i - u_{\beta})}{\Gamma(w_i - u_{\beta} - i\eta)}\frac{\Gamma(\bar{w}_j - u_{\beta})}{\Gamma(\bar{w}_j - u_{\beta} - i\eta)} \\\times\prod_{1 \le i < j \le M} \frac{(u_i - u_j)\, \Gamma(u_i - u_j - i\eta)}{\Gamma(u_i - u_j + i\eta + 1)} \ket{\Omega}.
\label{ansatzstatefinalformap}
\end{align}

\end{widetext}
\end{appendix}

\end{document}